\documentclass[12pt]{iopart}

\begin{document}

\title[Linking optical and IR obs. with gravitational wave sources through variability]{Linking optical and infrared observations with gravitational wave sources through variability}

\author{C W Stubbs}

\address{Department of Physics\\
Department of Astronomy\\
Harvard University\\
17 Oxford Street\\
Cambridge MA 02138 USA
}
\ead{stubbs@physics.harvard.edu}
\begin{abstract}
Optical and infrared observations have thus far detected more celestial
cataclysms than have been seen in gravity waves (GW). 
This argues that we should search
for gravity wave signatures that correspond to flux variability seen at 
optical wavelengths, at precisely known positions.  
There is an unknown
time delay between the optical and gravitational transient, but knowing the 
source location precisely specifies the corresponding time delays across the  gravitational antenna network as a function of the GW-to-optical arrival time difference. 
Optical searches should detect virtually all
supernovae that are plausible gravitational radiation sources. The
transient optical signature expected 
from merging compact objects is not as well understood, but there are good reasons
to expect detectable transient optical/IR emission from most of these sources as well. The next 
generation of deep wide-field surveys
(for example PanSTARRS and LSST) will be sensitive to subtle optical variability, but
we need to fill the ``blind spots'' that exist in the Galactic plane, and 
for optically bright transient sources.  
In particular, a Galactic plane variability survey at $\lambda \sim $2 $\mu m$ seems
worthwhile. 
Science 
would benefit from closer coordination between the various optical survey 
projects and the gravity wave community. 
\end{abstract}


\maketitle

\section{Introduction, and some assertions}

This paper addresses the pragmatic challenges of linking 
optical and infrared observations
with the time series recorded by the worldwide network of gravity wave antennae. 
This is particularly interesting as the optical astronomy community is moving 
towards an era of ``celestial cinematography'' where the entire accessible sky
will be imaged to unprecedented depth, with near-real-time detection and 
classification of variable sources. 

I will start with some assertions: 

\begin{enumerate}

\item{} More celestial transient and variable sources have been seen to date at 
optical wavelengths than have been seen in gravity waves. These 
detections include a diversity of supernovae, QSO's to very remote
distances, and the optical counterparts to GRB's. 

\item{} An uncoordinated patchwork of optical surveys are monitoring the 
sky at different revisit cadences, passbands, and flux levels. The 
next generation of surveys (such as PanSTARRS and LSST) will be sensitive to very subtle
changes in flux, for sources with variability timescales of more than a few days, across most
of the sky. We will soon start building an all-sky catalog of variability down to 22$^{nd}$ magnitude. 

\item{} In the context of potentially detectable gravity wave 
(GW) sources, optical monitoring 
currently has two ``blind spots'', one in flux and the other in 
sky coverage. A core collapse supernova in the local group
is (in my opinion) more likely to be first detected by eye than by a camera system; 
we are biased against detecting bright sources in our local neighbourhood, as they
would rapidly saturate the dynamic range of most survey systems. 
Similarly, the monitoring across the disk of our own Milky Way is incomplete. 
LSST, for example,
is considering avoiding the region with Galactic latitude $\mid b \mid < 20$, 
due to source confusion. 

\end{enumerate}

\section{Exploiting optical variability to trigger a constrained GW analysis}

These considerations suggest that (at least in the pre-LISA era) 
we should reverse the paradigm in which the 
gravitational wave observatories would, upon detection of a burst signal, 
pick up a red telephone and call the optical hotline, 1-(800)-CHASEIT.  
Instead, it seems that the expected sources of interest
have variability at optical wavelengths
that is more readily detectable than the associated gravity waves.  This 
approach can of course be combined with making optical followup observations
of regions on the sky that correspond to source directions from low-significance 
GW triggers. 

Using frame subtraction techniques ({\it e.g.} Barris {\it et al.}~2005, Miknaitis 
{\it et al.}~2007) 
the astronomical community
has achieved essentially Poisson-limited sensitivity to optical transients. 
Implementing pipelined image processing on cluster computers has 
allowed same-day detection and classification of optical transients. 

The main point of this paper is that we should exploit optical variability to search for gravity waves, in addition to the other way around. In a sense this amounts to 
making numerous ``pointed'' observations with the 
current generation of gravity wave detectors. 
I advocate a program in which {\it detected variability} of interest at optical and infrared
wavelengths be used as a prior for the analysis of the time series data from 
the network of gravity wave antennae. This approach can of course augment
and complement ``isotropic'' burst detection algorithms that ignore
known variable optical sources, but the advantage of searching for signals from a 
known direction is that the time delays and amplitude ratios of the detected
strain signals can presumably be used suppress the false detection rate.  
Furthermore, as was recently shown (LIGO \& Hurley (2007)) for a GRB source, even an upper bound
from the gravity wave detectors can be used to set limits, in SI units, on the
time rate of change of the mass quadrupole moment of the event. 
There is considerable 
merit in finding ways to exploit the upper limits to gravitational emission 
from celestial fireworks, at known distances, seen in the optical. 
Finally, one could imagine stacking, with appropriate time differences 
based on various source directions, the different antenna time series and {\it then} running a transient detection algorithm as a function of the time delay between the 
optical and gravitational transient. (I thank Bence Kocsis for making the
stacking suggestion.)

Making an association of GW sources with optical and IR data will greatly
facilitate both the fundamental physics exploitation and the 
astrophysical understanding of these systems. 
Another major goal in associating gravity wave sources with optical emission
is obtaining a GW-independent redshift of and (perhaps) distance to the source. This will 
be important in assessing systematics when constructing a Hubble
diagram with gravity wave sources (Holtz \& Hughes (2005); Deffayet \& Menou, (2007)).  

There are many stars and galaxies on the sky. By restricting our consideration for
GW linkage to those 
objects that exhibit appropriate {\it variability} and that lie within the likely 
GW detection range we can greatly reduce the 
surface density of candidate optical counterparts. This presumes that gravity
wave emission is frequently associated with optical variability. I argue below that this 
likely true. 

\subsection{An analysis strategy for seeking GW transients from a fixed location}

The positional uncertainties for optical variability are subarcsecond. This is 
to be compared with many arcminutes to degrees of 
expected positional uncertainty from 
even robust gravity wave detections (Sylvestre (2003)). 
This makes the optical-to-gravitational route considerably
easier than the converse. Knowing the position of the candidate GW source on the 
sky to an angular uncertainty $\sigma_{\phi}$ of an arcsecond specifies the time delays across the antenna network to a typical
precision of $\sigma_t \sim \sigma_{\phi} (B/c) cos(\phi(t))  $ where B is the baseline 
separation between the antennae, $\phi(t)$ is the 
angle between $\vec{B}$ and the direction to the source, and $c$ is the speed of
light. 
For the two LIGO detectors with a separation of 
$B=$3000 km this amounts to a typical uncertainty $\sigma_t$ of tens of nanoseconds.
The limited bandwidth of the GW antennae preclude taking full advantage of this, 
of course.

The GW search can be carried out by requiring joint excursions in strain at the 
various antennae in the network with 
a tightly constrained arrival time difference, but at an unknown arrival time (since we don't know the time delay between the optical and GW transients). The known location
specifies a unique relationship between the arrival time difference(s) and the 
time delay between the optical and the GW transient, as the Earth rotates relative
to the fixed source location.

\subsection{Optical variability from core collapse supernovae}

Core collapse supernovae do have intense transient optical
emission. Typical B-band absolute magnitudes for type II supernovae are
(Richardson {\it et al.} (2002))
M$_{B}^{SNII}\sim-18$. A detonation of such an object in M31, 
at a distance modulus of $\mu=24.3$,  would 
have a peak apparent magnitude of around $B=-18+24.3 \sim$ 6. 
By the standards of modern optical astronomy this is a {\it really} bright object!
A detonation in the Milky Way, even in heavily obscured regions near the 
Galactic center, would likely be even brighter. As discussed below, these
objects would saturate most current sky surveys, except at the very
beginning and end of their light curve. More exotic objects, such as
supernova 2006gy, which is thought to arise from the collapse of 
a very massive star (Smith {\it et al.}~2006), are even brighter in the optical.

We see more of these objects at optical wavelengths
than in gravity waves or neutrinos. Section 4 describes the numerous 
spectrally confirmed type II supernovae that were detected on the
sky during the S5 run.  

It seems safe to assume that any
core collapse SN that could produce a detectable gravitational signal would
have a huge, and readily detectable, variable optical signature. Since the 
current estimates ({\it e.g.} Ott et al. (2006)) for the strength of the SN gravity wave emission indicate 
LIGO only being able to see normal core collapse supernovae if they
occur within the local group, the challenge at optical wavelengths is 
fielding a system that would not immediately saturate.  
Given that the incidence of core collapse supernovae within the local group
is rare, it seems sensible to field the relatively inexpensive instrumentation 
that would generate precise early multiband light curves of such an event. 

\subsection{Optical variability from coalescing compact objects}

The mergers of compact objects are considered a likely source of 
detectable gravitational radiation. Ground-based interferometers and 
LISA have different spectral regimes
of optimal sensitivity, and therefore the compact object masses of interest
are different for the two detector types. 
A variety of mechanisms have been considered for producing 
electromagnetic emission in association with mergers. An illustrative 
sampling of these mechanisms includes:
\begin{itemize}
\item{} Optical emission powered by radioactive ejecta, as happens in SN Ia's (Rosswog (2005)), 
\item{} Conversion of gravitational radiation into electromagnetic waves through interactions in the local circumstellar environment (Sylvestre(2003)), and 
\item{} Gravitational radiation from black hole ringdown converting to electromagnetic radiation through local magnetic fields (Clarkson {\it et al.} (2004)).
\end{itemize}

An open question is what proportion of gravity-wave-detectable 
coalescence sources have undetectably low associated optical variability.
As we push the optical limits into the regime of 24$^{th}$ magnitude, I 
surmise that the optically silent fraction is low. Even in the case of BH-BH
mergers, the complex interaction between the emitted gravity waves,
the surrounding material, and any local magnetic fields could well 
produce detectable optical/IR emission. If we take the extreme case 
where the BH-BH events have no detectable associated optical transient, 
we'd only miss perhaps one third of the total merger events 
(see Table 2 of Belcyznski {\it et al.} (2002)) by triggering on optical emission.  

\subsection{The LIGO regime: mergers of 1 to 100 solar mass compact objects}

Sylvestre (2003) estimated the optical emission for compact object mergers 
in the LIGO band. Typical optical magnitudes for NS-NS mergers were
estimated as bring brighter than R = 20, even out to distances of hundreds
of Mpc. One mechanism described in that paper was expected to produce 
an optical transient that lasted about a day. This argues in favor of an optical 
wavelength sky survey that monitors the entire observable sky down to 
20$^{th}$ magnitude, daily. 

\subsection{The LISA regime: merging supermassive black holes}

A number of authors ({\it e.g.} Armitage \& Natarajan (2002); Milosavljevic \& Phinney (2006); Bogdanovic {\it et al.} (2007), Dotti {\it et al.} (2006))  
have explored the electromagnetic 
emission that might accompany the merger of compact objects
in the mass range (10$^5$-10$^7$ solar masses) that would 
produce GW signals in the spectral region where LISA is sensitive. 
These predictions for electromagnetic LISA counterparts exhibit considerable
variation. 

High mass BH-BH mergers (10$^5$-10$^7$ solar masses) are expected
to exhibit a substantial change in their Xray spectra, and some of this 
radiation may be re-processed into the OIR regime if the optical depths 
around the source are high enough. The gravitational and electromagnetic 
transients are expected to coincide within a $\Delta t$ of order a few years. 
Milosavljevic and Phinney (2006) also
refer to what they call a ``weak'' prompt electromagnetic transient that 
is simultaneous with the merger event, but they did not provide a 
quantitative estimate of this signature. 

Rosswog (2005) performed numerical simulations of NS-BH mergers
with small mass ratios, and found the ejection of substantial amounts of 
radioactive material that powers an electromagnetic transient comparable to that 
of a type Ia supernova. Armitage \& Natarajan (2002) considered coalescing objects
embedded in an accretion disk. 

There appears to be a rich phenomenology surrounding the merger
process for two supermassive black holes, and as yet no strong consensus 
on the likely optical/IR emission from such systems. 

Regarding the identification of optical counterparts of LISA sources, 
where we expect there to be numerous detections, a number of 
workers have considered the challenge of making the linkage in the GW-to-optical 
direction. For example
Kocsis {\it et al.} (2006) have considered the 3-d density of QSO's that might correspond to LISA sources.  

Given the attention that is already being given to 
this approach in the LISA era, the focus in this paper is making the connection
starting from optical detections, in the sparse-detection regime.  We therefore need
to review the various optical transient detection opportunities that exist and are
on the horizon, and the sky density of variable sources. 

\subsection{The Sky Density of Anomalous Variable Objects}

At R$\sim$21 there are a few thousand galaxies per square degree. The 
stellar density depends on Galactic latitude and longitude. If position on the sky is the only metric used to establish a linkage to an observed object, for GW positional uncertainties
of order a degree there is more than a thousand-fold ambiguity in identifying the optical 
counterpart. An estimate of GW sensitivity vs. distance can be used in conjunction
with photometric redshifts or some other estimator of distance
to reduce the catalog of potential host galaxies, but variability
provides important additional information.

Becker {\it et al.~}(2004) provide insights into the 
variable sky at faint flux levels, and that paper also provides a 
good overview of variability surveys. They conducted a search for faint transients. 
Their upper limit (95\% confidence) on variability on time scales of 1900 seconds
in the range 19.5$<$R$<$23.4 mag is 5.3 events deg$^{-2}$ day$^{-1}$. 
They detected three short duration transients in the B band, one of which is identified 
as a flare in a Galactic star. 

Data from the SDSS survey show (Sesar {\it el al.~}(2007)) 
that only about 2\% of the point sources on the sky with $g<$ 20.5 exhibit fractional flux variations of greater than 5\%. At $g\sim$ 20 quasars comprise
nearly two thirds of the photometrically variable sources. A combination of 
temporal variability, magnitude, and color information can do a good job of distinguishing 
between variable Galactic stars and extragalactic sources. 

The majority of the photometric variables can be readily classified into known 
bins of astronomical taxonomy. For our purposes this can be used to great advantage. 
For example, little GW emission is expected from RR Lyra and Cepheid 
variable stars. If we imagine that perhaps 1\% of the variables
exhibit anomalous behavior so as to be plausible transient GW counterparts, 
then their density the sky is a fraction of around
10$^{-4}$ of the static sources to R=20, or a few per square degree, 
which is a viable regime. 
Whether optical transients trigger GW searches, or whether the goal is to go in 
the other direction and 
associate optical sources with GW triggers, concentrating on 
anomalous variable sources will ease the task. 

\section{A diversity of optical survey systems}

There are a number of options when scanning the skies for optical variability 
that might have associated gravitational wave emission. For a fixed 
detector area, in square cm, the focal length of the optical system 
determines the field of view (a solid angle $\Omega$). There is a 
practical limit to how fast the optical system can be, and so for a fixed 
detector area a larger aperture optical system 
necessarily has a smaller angular field of view. The etendue of the overall system, 
the $ \pi R^2 \Omega$ product of collecting area times field of view, 
is an important figure of merit for a survey system. The faintest source that can 
be detected, however, depends only on the collecting area and not on the 
field of view. Table 1 presents the relevant properties of some representative optical 
survey systems, and illustrative examples are described in the subsections
that follow.  

\subsection{All-sky cameras}
At one extreme are instruments that use a fisheye lens to 
monitor the entire accessible sky at once. Due to the optical constraints 
described earlier, these systems have very small effective apertures. A nice
example of implementing this approach is the ConCam system 
(Shamir \& Nemiroff (2005)), 
which has now been installed at multiple observatories. The sensitivity of these
systems is comparable to that of the naked eye, around 6$^{th}$ magnitude, 
and they are usually pointed at the zenith and left to run unattended.  

\subsection{All-sky surveys}

A system optimized for reaching 
fainter magnitudes, but also with full sky coverage, is the All Sky Automated Survey (ASAS) 
(Paczynki, 2006). ASAS was designed to detect variability across the entire
sky, down to 14$^{th}$ magnitude, capturing an image of each patch of sky
with daily revisits.  The field of view of ASAS allows a succession of images to cover
the accessible sky once per night. An ambitious set of next-generation 
sky surveys, yet to enter full operation, are discussed below.  

\subsection{GRB followup systems}

A  suite of instruments was designed to make rapid optical followup observations
of GRB's. This includes the ROTSE series of devices (Yost {\it et al.~}(2006)), the WASP instrument (Pollacco {\it et al.~}(2006)),
and Raptor (Wozniak {\it et al.~}(2005)). The field of view of these systems was initially chosen to 
correspond to the expected location uncertainty from BATSE, 
around 15 degrees, but subsequent improvements in GRB locations have led
to smaller fields and correspondingly larger apertures. Some of these groups
are now contemplating or are undertaking sky surveys with these instruments. 
 
\subsection{Killer asteroid, planetary transit, and microlensing surveys}

These are time domain surveys, usually over a limited 
portion of the sky. 
Surveys designed to detect potentially hazardous asteroids are typically
implemented on 1 m class telescopes, in order to attain sensitivity to 
asteroids smaller than 1 km in size. This reduces the field of view, and 
so these systems usually concentrate their observations in the plane of the 
ecliptic, often avoiding regions of low Galactic latitude. One of the more successful
such systems is LINEAR (Stokes {\it et al.~}(1998)). The asteroid surveys typically take re-observations of a field about a half hour after the night's initial image. 

Transit surveys take multiple images of fields centered on relatively bright (to allow
high resolution spectroscopic radial velocity followup) stars in order
to search for the photometric signature of a transiting planet. While these
projects reach impressive levels of sensitivity to flux changes from the target
star, their sky coverage is typically insufficient to be of interest in the context of 
gravity wave counterparts. 

Another example of time domain systems that survey a limited region on the 
sky are the microlensing search systems. Of the telescope systems that were 
originally specifically designed to carry out microlensing surveys, one (MACHO) was lost to an Australian brushfire, one (EROS) is no longer
in operation, and one (OGLE) continues to monitor fields in the Magellanic Clouds
and fields in the Galactic plane.  

\subsection{Dedicated supernova searches}

A relatively small field of view (perhaps 10 arcminutes across) can be
used to monitor specific nearby galaxies. This is the path taken by
the KAIT supernova system (Filippenko {\it et al.~}(2001)), sited at Lick Observatory. 
This telescope 
runs through a list of target galaxies, and the successive images are 
searched for the transient brightening of a point source that is characteristic
of a supernova. 

\subsection{Multipurpose telescopes}

The Sloan Digital Sky Survey (SDSS) has both an imaging and a spectroscopic
mode. A portion of the time on the SDSS telescope has been used to do repeated imaging
(Sessar {\it et al.~}(2007)) on one particular region, SDSS stripe 82. 
SDSS is an example of a system
that has good area coverage capability, but other scientific priorities 
(including spectroscopy) limit
the time spent on time domain imaging, such that high cadence all-sky
imaging is not currently being undertaken. Other telescopes with wide field cameras
({\it e.g.} CFHT, SUBARU, and the NOAO 4 meter telescopes at Kitt Peak and CTIO) are allocated
to a diversity of projects, and this limits their spending a large fraction of the telescope
time for a multi-epoch sky survey.  

\subsection{The next generation: PanSTARRS, SkyMapper, VST, and the LSST}

The astronomical community is now building multiple wide field sky survey
systems. These include PanSTARRS, on the Haleakala volcano on 
the island of Maui, SkyMapper, 
destined for the Siding Spring Observatory in Australia, the VLT Survey Telescope 
(VST) in Chile, and the Large Synoptic Survey Telescope (LSST), also 
destined for construction in Chile. PanSTARRS-1, VST, and SkyMapper are all 
slated to 
start science observations sometime in 2008. Of particular interest
in the context of gravitational wave sources, the PanSTARRS-1 
system has an auxiliary imager, the Imaging Sky Probe (ISP), that uses
a single imager and a small aperture fore-optic that is co-boresighted with the 
main instrument to extend the bright 
end of the dynamic range up to brighter than $\sim$8$^{th}$ magnitude. This 
ISP will be operated in tandem with the main imager, mainly for calibration 
purposes.
The VST system is not dedicated
to a single integrated survey strategy. The LSST
is the most ambitious of these systems, and current plans call for it to begin operation
in 2014. Both LSST and the multi-aperture version of PanSTARRS (PS-4) are 
currently ``unencumbered by full funding''. 

While these systems will jointly provide variability detection down to very faint
flux levels, the different projects have different goals for sky coverage. For example
at the present time the LSST default plan avoids observing in regions of 
low Galactic latitude, due to source confusion. It is important to recognize
that as these new survey systems come online, what is now rare will become
commonplace, and we will catch glimpses of processes that are
presently unobservably infrequent.

The PanSTARRS-1 telescope is expected to detect hundreds of supernovae
{\it per month}, and dozens of these are expected to be brighter than R$=$20.
Among the imminent next-generation systems, PanSTARRS-1 (PS-1) is unique in its
goal of real-time detection of transient phenomena at faint levels, across the 
entire observable sky. In particular PS-1 is well suited
to detecting faint optical ``orphan afterglows'' that might accompany compact 
object mergers, but that beam their gamma rays away from us. Zou {\it et al.} (2007) 
estimate that a typical afterglow should be brighter than R=21 for many days. 
According to their Figure 3 we might expect PS-1 (which revisits each 
patch of 3$\pi$ sr often enough to not miss one) to see one such source at any 
given time. 

\begin{table}[htdp]
\caption{Some Optical Sky Surveys. This is a non-exhaustive list of existing
and planned ($^*$) optical sky surveys. One figure of merit for a survey 
is the product of collecting area times its field of view (FOV), 
the A$\Omega$ product. Another figure of merit is the 
source intensity to which it is sensitive, which is primarily a function of 
only aperture.  The limiting point source magnitudes listed in the final column
are very approximate, meant to illustrate the rough range of sensitivities. 
Similarly, the A$\Omega$ products ignore obscuration in the optics, 
but do give a rough basis for comparison between systems. An interesting
comparison is that the Hubble Space Telescope has A$\Omega=$0.016 (mdeg)$^2$.}
\begin{center}
\begin{tabular}{lccccccc}
\hline
System &	Diam. &	FOV  &	A$\Omega$  & Mode & $\sim$ ``Sensitivity'' \\
~ & (m) & (deg) & (m deg)$^2$ & ~ & ~60 sec 5 $\sigma$\\
\hline
ConCam	& 0.004	&180 & 0.4 & All sky continuous & 6 \\
WASP &	0.1 &	15	& 1.7 &	Triggered \& Partial Survey & 15 \\
ROTSE-III &	0.45 &	2	&0.6& 	Triggered \& Partial survey & 18 \\
Raptor &	0.07	& 35 &	4.7	& Triggered & 16 \\
ASAS  &	0.10	& 3	& 0.07	& All sky, daily & 14 \\
HAT & 0.1 & 8 & 0.5 & Precision partial survey & 15\\
LINEAR &	1.0	& 1.4	& 1.5 & 	Ecliptic & 19 \\
SDSS &	2.5	& 1.5 	& 11	& Partial survey & 22 \\
Palomar/QUEST & 1.2 & 4 & 18 & Multicolor partial survey & 22 \\
$^*$VST &	2.6	& 1	& 5.3	& Allocated time & 22 \\
$^{*/2}$PS-1 &	1.8	& 2.6	& 23	&All-sky Survey & 22 \\
$^*$PS-4 & four x 1.8 & 3 ea & 91 & All-sky survey & 23 \\
$^*$LSST&	8.5	& 3 	& 510 &	``All-sky'' Survey & 24 \\
KAIT  &	0.8	& 0.1 &	0.005	& Nearby galaxies & 19 \\
\end{tabular}
\end{center}
\label{default}
\end{table}

\subsection{What about the infrared?}

The discussion here emphasizes optical surveys, taken to be the 
wavelength region over which silicon CCDs are sensitive, longward
of the ozone cutoff:  350 nm $ < \lambda< $ 1000 nm. There is merit
in considering wide-field surveys at longer wavelengths, particularly 
towards the obscured regions near the Galactic center. Lucas {\it et al.} (2007)
describe the UKIDSS Galactic Plane Survey in the J,H,K bands, but 
I am not aware of a program to do variability monitoring using the IR
across the Galactic plane.  It is well worth considering
undertaking an IR Galactic variability survey to augment the 
gravitational wave searches. 
If the coalescing compact object mergers are enshrouded in and/or
behind 
optically thick material, IR variability would be a useful way to 
identify potential GW sources even through heavy attenuation in the optical. 

A typical core collapse SN has an IR peak magnitude of $K\sim -18$ (Grossan {\it et al.} (1999)). 
Table 2 illustrates the stunning difference in transmission through ``dust''
in the Galactic center for light in the K band ($\lambda=$2.2 $\mu$m) vs. light in the V band (0.55 $\mu$m). 

\begin{table}[htdp]
\caption{A comparison of extinction effects in the V and K bands. 
The columns list extinction (in magnitudes) in V and K in the 
first two columns, assuming a canonical relationship of $A_K=0.112*A_V$. 
The third and fourth columns present the expected
peak magnitude for a heavily extincted type II supernova at a distance of 10 kpc,
essentially on the other side of the Galaxy. Working thorough a region 
with $A_V>20$ in the optical is very hard, while in the K band it is entirely
tractable to imagine detecting a typical type II SN through 150 magnitudes of 
$V$ band attenuation, using a 1 meter class telescope!  }
\begin{center}
\begin{tabular}{rrrr}
$A_V$  & $A_K$ &  $V$ & $K$ \\ 
\hline
0	& 0	&	-3 &	-3\\
5 &	0.6 &	2 &	-2.4\\
10 &	1.1 &	7 &	-1.9\\
15 &	1.7 &	12 &	-1.3\\
20 &	2.2 &	17 &	-0.8\\
50 &	5.5 &		47 &	2.6\\
100 &	11.1 &	97 &	8.1\\
150 &	16.7 &	147 &	13.7\\
\end{tabular}
\end{center}
\label{default}
\end{table}

\section{Existing optical detections are already interesting}

Of the supernova detections listed on the CfA web site 
\linebreak
(http://cfa-www.harvard.edu/iau/lists/RecentSupernovae.html) 
I found 762 that coincided with the S5 science run of LIGO and VIRGO. 
Of these, 411 were spectroscopically confirmed as
core collapse (type II) SNe, and 89 of these were in catalogued host galaxies. 
I then used SIMBAD to query the list of named galaxies and obtain their catalogued
redshifts. Using an estimate of H$_o$=75 km/sec per Mpc I generated an 
approximate distance to each of these supernovae.
The closest of the type II SNe from S5, in named host galaxies, are listed in Table 3.
 
\begin{table}[htdp]
\caption{Three relatively close type II supernovae that coincided with LIGO run S5. 
The columns
list the IAUC supernova designation, its host galaxy, and an estimate
of the distance.The progenitor characteristics are from Li {\it et al.} (2007). Since these
were drawn from only 89 of the 762 SNe that occurred during run S5, there
are likely others that were closer.}
\begin{center}
\begin{tabular}{llll}
IAUC SN & Host & D (Mpc) & Progenitor notes \\ 
\hline
2006bp & NGC 3953 & 14 &  \\
2006my & NGC 4651 & 23 & 7-15 solar mass star \\
2006ov & M61 & 21 & 12-20 solar mass star \\
\end{tabular}
\end{center}
\label{default}
\end{table}

While the current estimates of gravitational radiation from these particular objects
do not lead us to expect a detectable signal, it certainly makes sense to look. 
In fact, it seems to me a good idea to obtain a distance estimate
to every galaxy that hosted a supernova during run S5, and to search for
evidence of strain signals that might be associated with these events. The host
galaxy redshifts can often be extracted from the same spectra used to determine
the SN type. 

\section{Blind Spot \#1: The Milky Way plane}

Except for the all-sky camera systems described above and some of the 
microlensing projects, most  
sky survey programs intentionally avoid observing the plane of the Milky Way.
This means that for sources fainter than about 6$^{th}$ magnitude, we 
are currently blind to transient sources that lie close to the Galactic plane! 
This is of course precisely where we'd expect to find supernovae in our own
Galaxy. The historical supernovae in the Milky Way occurred (Richmond 2007) at Galactic latitudes of b$= 0.1, 0.8, 32, -6.5, 2.7, 1.2 $ and $7.6.$ degrees. 
Now that we realize objects like the Sagittarius galaxy lie behind the Galactic center,
a deeper Galactic plane monitoring program seems  well motivated. 
We would also presently miss transient events in external galaxies that
lie in the ``zone of avoidance'' behind the Galactic disk. I re-iterate that 
frame subtraction analysis can be used to suppress the light from static 
sources. If we focus on variability detection then the deleterious effects of 
looking through the Galactic disk are added sky background, and 
extinction. Augmenting the all-sky optical cameras like ConCam with a wide field imager
at $\lambda \sim$ 2$ \mu m$ would be very inexpensive. 

\section{Blind Spot \#2: Really bright transients}

The extreme disparity between the optical and gravitational signal strenghts 
suggests that we pay attention to being able to measure (in multiple filters) the transient 
OIR flux from bright sources. For a bright core collapse supernova
within the local group, the professional optical and IR astronomy communities are not in a position to exploit well-calibrated instruments to observe a naked eye object. 
As the frontier of optical astronomy has shifted to ever fainter sources, smaller
telescopes have been relegated to lower priority. The SN 1987A experience suggests
that the astronomical community will adapt, if necessary with neutral density
filters duct taped into the instrument, but the time lost to scrambling will lose
precious early light curve data.  Local supernovae are very rare, and we are
not currently prepared to detect and follow one. It's the prospect of detecting optical and  gravitational radiation plus neutrinos that makes these bright objects so special, and we should prepare accordingly.

\section{Opportunities, and suggested next steps}

A summary of the opportunities and obvious next steps includes:
\begin{itemize}

\item[] {- Implement real-time frame subtraction analysis of all-sky camera data.} 

\item[] {- Assess merits of optical vs. IR Galactic plane monitoring program.}

\item[] {- Build and operate optimal Galactic plane monitoring system.} 

\item[] {- Calculate optical and IR variability that accompanies coalescence events.} 

\item[] { - Assess and if needed augment optical all-sky monitoring projects.}

\item[] {- Search for gravitational wave signals associated with all known SNe, 
and all unclassified optical transients.} 

\item[] {- Obtain distances and estimates of $t_0$  for all SNe that coincide with run S5.}

\item[] {- Establish stronger collaborations across the optical, IR, gravitational and neutrino communities.}

\end{itemize}

\section{Acknowledgments, and a dedication}

I am grateful to the organizers of the 12$^{th}$ Gravitational Wave Data Analysis
Workshop, held in Cambridge MA in Dec 2007, for the invitation to attend and to present this perspective. My interest 
in this topic was re-kindled during a visit to Syracuse University, and I thank 
Duncan Brown and Peter Saulson for interesting conversations during that trip.
It was a pleasure to speak with a long-standing friend, Carl Akerlof, about
the ROTSE system. 
I also am grateful for conversations with Avi Loeb and Bence Kocsis, at the 
Center for Astrophysics at Harvard. I especially thank B. Kocsis and D. Brown 
for their helpful comments on an early draft of this manuscript. 
Cullen Blake brought the LIGO/GRB preprint by to my attention. 
I dedicate this paper to the memory of 
Bodhan Paczynski, who was a tireless advocate for continual optical monitoring 
of the sky. 
 
\section*{References}

\smallskip
\begin{harvard}
\item[] Armitage, P J \& Natarajan, P 2002 {\it ApJ} {\bf 567}, L9.  
\end{harvard}
\smallskip

\smallskip
\begin{harvard}
\item[] Barris, B., Tonry, J.L., Novicki, M.C., Wood-Vasey, W.M 2005 {\it AJ} {\bf 130}, 
2272. 
\end{harvard}
\smallskip

\smallskip
\begin{harvard}
\item[] Becker, A.C.~{\it et al.~} 2004 {\it ApJ} {\bf 611}, 418.
\end{harvard}
\smallskip

\smallskip
\begin{harvard}
\item[] Belczynski, K {\it et al.} 2002, {\it ApJ} {\bf 572}, 407.
\end{harvard}
\smallskip

\smallskip
\begin{harvard}
\item[] Bogdanovic, T, Smith, B D, Siguresson, S and Eracleous, M 2007 {\it arXiv:0708.0414}, to appear in {\it ApJ Supp}.
\end{harvard}
\smallskip

\smallskip
\begin{harvard}
\item[] Clarkson, C A {\it et al.} 2003 {\it ApJ} {\bf 613}, 492
\end{harvard}
\smallskip

\smallskip
\begin{harvard}
\item[] Deffayet, C and Menou, K {\it et al.} 2007  arXiv:0709.0003 
\end{harvard}
\smallskip

\smallskip
\begin{harvard}
\item[] Dotti, M {\it et al.} 2006 {\it MNRAS} {\bf 372}, 869. 
\end{harvard}
\smallskip

\smallskip
\begin{harvard}
\item[] Filippenko, A.~{\it et al.~} 2001 {\it ASP Conference Series} {\bf 246} 121
\end{harvard}

\smallskip
\begin{harvard}
\item[] Grossan, B {\it et al.} 1999 {\it AJ}, {\bf 118}, 705
\end{harvard}
\smallskip

\smallskip
\begin{harvard}
\item[] Holz, D \& Hughes, S A 2005 {\it ApJ} {\bf 629}, 15.  
\end{harvard}
\smallskip

\smallskip
\begin{harvard}
\item[] Kocsis, B {\it et al.} 2006 {\it ApJ} {\bf 637}, 27  
\end{harvard}
\smallskip


\smallskip
\begin{harvard}
\item[] LIGO Collaboration \& Hurley, K.  2007, arXiv:0711.1163
\end{harvard}
\smallskip

\smallskip
\begin{harvard}
\item[] Lucas, P W {\it et al.} 2007 {\it MNRAS (submitted)}, arXiv:0712.0100  
\end{harvard}
\smallskip

\smallskip
\begin{harvard}
\item[] Milosavljevic, M \& Phinney, S 2006 {\it ApJ} {\bf 622}, L93.  
\end{harvard}
\smallskip

\smallskip
\begin{harvard}
\item[] Ott, C {\it et al.} 2006 {\it PRL} {\bf 96}, 1102.  
\end{harvard}
\smallskip

\smallskip
\begin{harvard}
\item[] Paczynski, B 2006 {\it PASP} {\bf 118}, 1621
\end{harvard}
\smallskip

\smallskip
\begin{harvard}
\item[] Shamir, L and Nemiroff, R.J.~2005 {\it PASP} {\bf 117}, 972.
\end{harvard}
\smallskip

\smallskip
\begin{harvard}
\item[] Pollacco D K {\it et al.} 2006 {\it PASP} {\bf 118}, 1407
\end{harvard}
\smallskip

\smallskip
\begin{harvard}
\item[] Richardson, D {\it et al.} 2002 {\it AJ}, {\bf 123}, 745.
\end{harvard}
\smallskip

\smallskip
\begin{harvard}
\item[]Richmond, M. 2007, www.tass-survey.org/richmond/answers/historical.html
\end{harvard}
\smallskip

\smallskip
\begin{harvard}
\item[] Rosswog, S 2005 {\it ApJ} {\bf 634}, 1202
\end{harvard}
\smallskip

\smallskip
\begin{harvard}
\item[] Sessar, B.~  {\it et al.} 2007 {\it AJ}, {\bf 134}, 2236.
\end{harvard}
\smallskip

\smallskip
\begin{harvard}
\item[] Smith, N {\it et al.} 2007  {\it ApJ} {\bf 666}, 1116
\end{harvard}
\smallskip

\smallskip
\begin{harvard}
\item[] Stokes, G.H.~{\it et al.~} 1998 {\it BAAS} {\bf 30}, 1042. 
\end{harvard}
\smallskip

\smallskip
\begin{harvard}
\item[]Sylvestre, J 2003 {\it ApJ}, {\bf 591}, 1152
\end{harvard}
\smallskip

\smallskip
\begin{harvard}
\item[] Miknaitis, G.~{\it et al.} (the ESSENCE Collaboration) 2007 {\it ApJ} {\bf 666}, 674.
\end{harvard}
\smallskip

\smallskip
\begin{harvard}
\item[] Wozniak, P.R.~{\it et al.~}2005 {\it ApJ} {\bf 627}, L13.
\end{harvard}
\smallskip

\smallskip
\begin{harvard}
\item[] Yost, S.A.~{\it et al.~}2006 {\it Astronomische Nachrichten} {\bf 327}, 803
\end{harvard}
\smallskip

\smallskip
\begin{harvard}
\item[]
\end{harvard}
\smallskip

\end{document}